\documentclass[pdflatex,sn-nature]{sn-jnl}

\usepackage{graphicx}%
\usepackage{multirow}%
\usepackage{amsmath,amssymb,amsfonts}%
\usepackage{amsthm}%
\usepackage{mathrsfs}%
\usepackage[title]{appendix}%
\usepackage{xcolor}%
\usepackage{textcomp}%
\usepackage{manyfoot}%
\usepackage{booktabs}%
\usepackage{algorithm}%
\usepackage{algorithmicx}%
\usepackage{algpseudocode}%
\usepackage{listings}%

\usepackage{gensymb} %for degree symbol

\theoremstyle{thmstyleone}%
\theoremstyle{thmstyletwo}%

\theoremstyle{thmstylethree}%

\makeatletter
\renewcommand\@biblabel[1]{#1.}
\makeatother

\begin{document}

\title[Article Title]{\bf{Programmable Persistent Random Walks in Active Brownian Particles Govern Emergent Dynamics}}

\author[1]{\fnm{Tarun} \sur{Sunkesula Raghavendra}}
\author[1]{\fnm{Yogesh} \sur{Shelke}}
\author[1]{\fnm{Stijn} \sur{van der Ham}}
\author[1]{\fnm{Anpuj} \sur{Nair S}}
\author*[1]{\fnm{Hanumantha Rao} \sur{Vutukuri}}\email{h.r.vutukuri@utwente.nl}
\affil[1]{\orgdiv{Active Soft Matter and Bio-inspired Materials Lab, Faculty of Science and Technology, MESA+ Institute for Nanotechnology and Center for Brain-Inspired Nano Systems}, \orgname{University of Twente}, \orgaddress{\city{Enschede}, \postcode{7500 AE}, \country{The Netherlands}}}

\maketitle

\begin{abstract}
\noindent 
{\bf {
Self-propelled particles serve as minimal models for emulating the dynamic self-organization of microorganisms, yet most synthetic systems remain limited to a single mode of motion, namely active Brownian particles (ABPs). Here, we present an experimental strategy to encode various persistent random walks in ABPs by combining light-modulated propulsion strength with magnetic control of propulsion direction. Our system enables programmable Lévy walks with tunable step-length distributions, run-and-tumble dynamics, self-avoiding random walks, and Gaussian walks, with on-demand switching between motion modes within a single experiment. In addition, particles are steered along complex trajectories such as Fibonacci spirals and nested polygons. Beyond single-particle behavior, we show that propulsion modes influence clustering dynamics by comparing ABPs with chiral active particles undergoing circular motion. These results establish a versatile platform for investigating how encoded motion at the level of individual particles governs transport, search strategies, and emergent organization in active matter systems.}}\\

\end{abstract}
{\bf{Keywords:}} Active matter, Self-organization, Persistent random walks, L\'evy walks, Chiral clusters

\section*{Introduction}\label{sec1}
Hallmark features of living systems, such as autonomous motion, adaptability, evolution, defense mechanisms, and effective food search, are essential for survival and functionality. These unique features have fueled the emergence of active matter within the realm of soft matter physics. Micron-sized synthetic self-propelled particles (SPPs), modeled as active Brownian particles (ABPs), uniquely convert chemical energy into directed mechanical motion, offering a powerful experimental, theoretical, and simulation platform for understanding the collective behaviour of inherently out-of-equilibrium systems \cite{bechingeractive, gompper_review2025, zottl-review, palacci18, vutukuri2020active}. 

However, despite these capabilities, a key limitation of the ABP model is its inability to replicate the more complex and highly efficient motility strategies observed in living organisms. For instance,  wild-type rod-shaped \textit{Escherichia coli} (\textit{E. coli})  exhibit run-and-tumble (RTP) motion, characterized by alternating straight runs and random tumbles that reorient the bacterium. Many studies have reported that these run lengths typically follow an exponential distribution, resulting in long-time diffusive behavior reminiscent of ABPs \cite{berg2004coli,lovely1975statistical,villa2020run,Rao_side_propelled}. However, under certain conditions, \textit{E. coli} can switch to Lévy walk dynamics, where run lengths follow a power-law distribution, driven by chemotaxis noise and resulting in anomalous transport \cite{huo2021swimming,ariel2015swarming}. This adaptive search behavior extends beyond microorganisms; for example, larger organisms, such as seabirds during foraging or predators while hunting, employ sophisticated strategies like Lévy walks to efficiently search vast environments with limited information \cite{viswanathan1996levy, humphries2012foraging, reynolds2009honeybees}. These Lévy walks are modeled as random walks with power-law-distributed run-lengths, leading to superdiffusive behavior, which provides a significant advantage when target locations are unknown. 
 
While such strategies have been successfully implemented in macroscopic inanimate systems, such as robotic platforms \cite{Termite, rubenstein2014programmable}, achieving comparable control at the micron scale remains elusive and poses a major challenge with profound implications for advancing active matter research.
The main limitation in achieving such adaptive search strategies lies in the inherent rotational diffusion of ABPs. While this model captures the fundamental aspects of active motion \cite{bechingeractive,Howse_SMP_2007, vutukuri2020light}, it cannot replicate the complex and adaptive persistent random walks that living organisms have evolved for survival and exploration. Moreover, search strategies can be influenced by the topology of the environment \cite{volpe2017topography,bechingeractive}. Several studies have utilized light or magnetic fields to steer particles \cite{shen2023magnetically,demirors2017colloidal,yousefi2021independent,mano2017optimal,al2020magnetically,zehavi2025programmable,pellicciotta2023light,fernandez2020feedback,tiernoAutonomouslyMovingCatalytic2010,klineCatalyticNanomotorsRemoteControlled2005}, optical control visual-perception-based strategies \cite{lavergne2019group,muinos2021reinforcement}, and complex shape-reconfigurable particles \cite{lucio-pnas} to control and program their paths. However, the ability to encode diverse persistent random walks in synthetic self-propelled microparticles remains a significant challenge and is largely elusive. 

Here, we introduce a simple yet versatile experimental strategy for precise control of particle motion, enabling the encoding of various persistent random walks, namely Lévy walks, RTP motion, Gaussian walks, and self-avoiding walks (SAWs), into ABPs. Using magnetically responsive particles, we control propulsion direction with an external magnetic field programmed via an Arduino and modulate propulsion strength through UV-light intensity. This approach enables rapid, on-demand switching between motion modes within a single experiment, allowing the creation of programmable velocity landscapes that mimic non-uniform environments. At the single-particle level, distinct walk statistics and run-length distributions determine the scaling of the mean-squared displacement (MSD) and end-to-end distance. We further steer particles along prescribed trajectories, from polygonal paths to nested patterns and Fibonacci spirals. Beyond single-particle dynamics, distinct propulsion modes influence clustering behavior, leading to differences in cluster growth between ABPs and particles undergoing circular motion.

\section*{Results and Discussion}
\label{sec:Results_and_discussion}

Our system consists of a hematite cube of size 0.9 \textmu m embedded in a 3-(trimethoxysilyl)propyl methacrylate (TPM) particle, 1.5 \textmu m in size, as shown in the scanning electron microscope (SEM) image in the inset of Fig.~\ref{fig:Pareto}a (see also Supplementary Note 1). Under UV (365 nm) illumination in the presence of fuel (6\%\,v/v hydrogen peroxide, $\mathrm{H}_2\mathrm{O}_2$), these particles exhibit self-propulsion via diffusiophoresis \cite{zhou2021ionic, palacci18, li2022programmable} (for details, see Methods). By leveraging both light and magnetic fields, we achieve tunable propulsion strength through light intensity modulation, while the propulsion direction is precisely controlled via an external magnetic field, which exploits the weak ferromagnetic nature of the hematite cube to regulate particle orientation \cite{palacci18, meijer2021}. Our experimental setup is described in Supplementary Note 2.

\subsection*{Lévy walks with varying power-law exponents}
\label{subsec:levy walks}

We first demonstrate how to encode power-law distributions with different exponents in run-lengths or persistence lengths, allowing us to explore how different run-length distributions affect the MSD and the nature of particle transport in ABPs over timescales longer than their reorientation time. This potential anomalous diffusion can be modeled using Pareto distributions, a family of heavy-tailed distributions following a power law. In contrast, classical random walks, such as those in typical Brownian motion, exhibit run lengths that follow a Gaussian distribution with light tails. For a particle moving with constant speed $v$, akin to ABPs, the run-lengths $\ell=v\tau_\mathrm{R}$ follow the same scaling as the run-times $\tau_\mathrm{R}$. The Pareto distribution for the run-time $\tau_\mathrm{R}$ is given by
\begin{equation}
    P(\tau_\mathrm{R}) = \frac{\tau_0^{\gamma}\gamma}{\tau_\mathrm{R}^{1+\gamma}}\,\text{,} \quad \quad \text{if } \quad \tau_\mathrm{R}\geq\tau_0\text{,}
    \label{eq:pareto_dist}
\end{equation}

\noindent
where $\gamma$ is the scaling exponent that determines how quickly the tail decays and $\tau_0$ is the lower cutoff for $\tau_\mathrm{R}$, such that $P(\tau_\mathrm{R})=0$ if $\tau_\mathrm{R}<\tau_0$ \cite{zaburdaev2015levy}. The mean run-time is given by $\langle \tau_\mathrm{R} \rangle = \tau_0\gamma / (\gamma-1)$ for $\gamma>1$ \cite{karani2019tuning}. In our experiments, we set $\tau_0=2$\,s. The MSD of the particle executing a random walk with run-times distributed according to Equation \eqref{eq:pareto_dist} scales as follows,

\begin{equation}
    \langle r^2(t) \rangle = t^\alpha =
    \begin{cases}
        \begin{array}{r l}
        t^2, & \quad 0<\gamma<1\\
        t^{3-\gamma}, & \quad 1<\gamma<2 \,\text{,}\\
        t, & \quad \gamma>2
        \end{array}
    \end{cases}
    \label{eq:MSD_pareto}
\end{equation}

\noindent
where $\alpha$ is the scaling exponent that determines the type of diffusion behavior \cite{zaburdaev2015levy}. For lower $\gamma$ values, the run-time distribution has a heavier tail, indicating a higher probability of longer runs, which is reflected in the MSD as a ballistic scaling at longer times. In contrast, higher $\gamma$ values correspond to shorter runs, resulting in more diffusive scaling. Therefore, the superdiffusive regime is characterized by an intermediate range $1 < \gamma < 2$, which defines the relation between $\gamma$ and $\alpha$ as $\gamma + \alpha = 3$.

We first selected two different power-law exponents, inspired by the persistent random walks that living systems use to explore their phase space efficiently. First, $\gamma = 1.2$ was motivated by wild-type \textit{E.\,coli}, which navigate their environment using Lévy walks driven by chemotaxis signaling noise \cite{huo2021swimming}. Second, $\gamma = 1.5$ represents the classic Lévy walk, an efficient search strategy observed in seabirds during foraging and predators during hunting \cite{zaburdaev2015levy}. To further demonstrate the versatility of our approach, we also used $\gamma = 1.8$ \cite{karani2019tuning}, expanding the range of behaviors and showcasing the adaptability of our system with three distinct $\gamma$ values in our experiments.

\begin{figure}[H]
    \centering
    \includegraphics[width=1\linewidth]{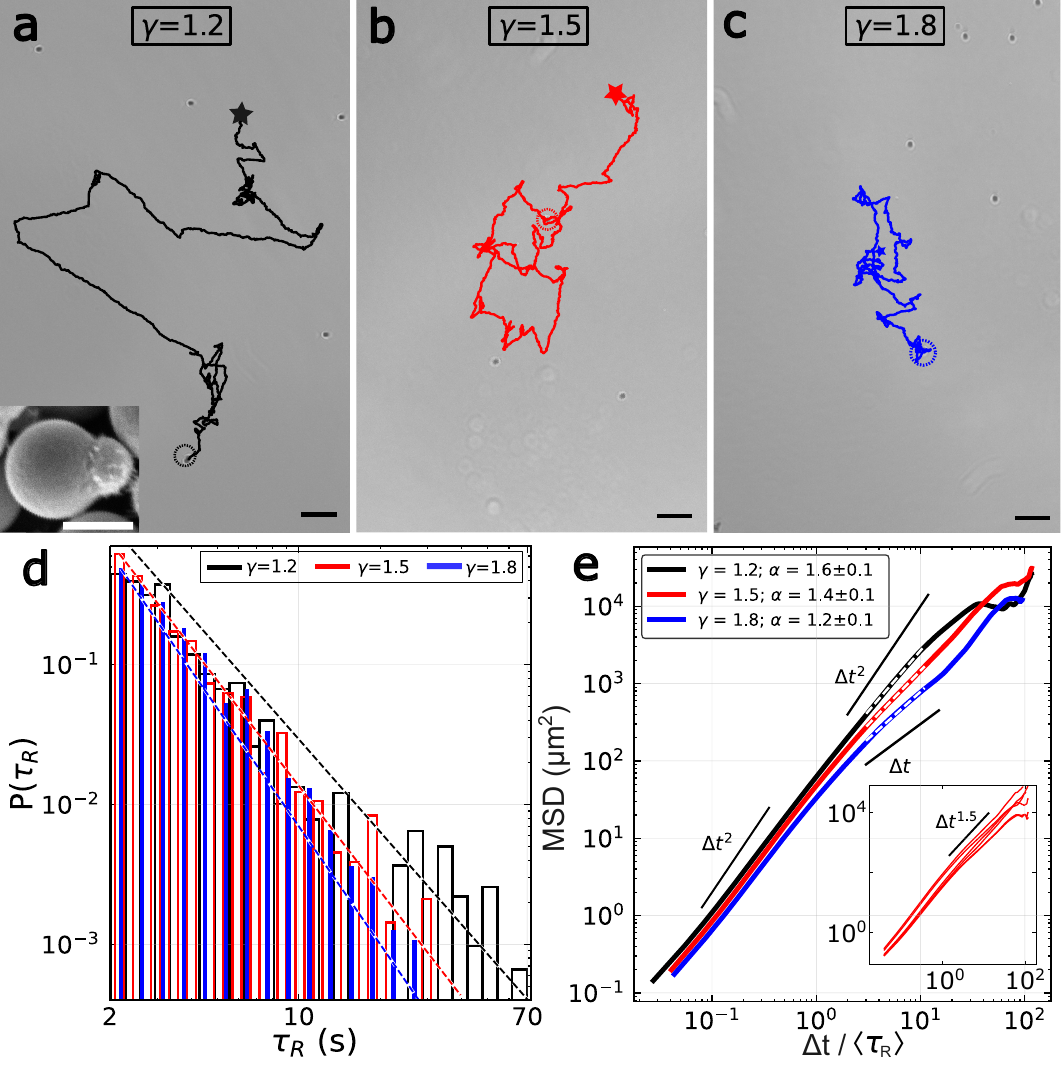}
    \caption{\textbf{L\'evy walks with three different power law distributions.} 
    \textbf{a-c)} Overlaid particle trajectories on bright-field microscope images showing different run-lengths for $\gamma = 1.2$, $1.5$, and $1.8$, respectively. The exponent $\gamma$ determines the decay of the run-time distribution (see Equation \eqref{eq:pareto_dist}). The star and circle represent the start and end positions of the particle, respectively. Inset: SEM image of the particle. \textbf{d)} Distributions of run-times ($\tau_\mathrm{R}$) for different $\gamma$ values, on a log-log plot highlighting the long power-law tails. The dashed lines represent the theoretical Pareto distributions from which the run times were taken, with respective exponent values $\tau_\mathrm{R}^{-(1+\gamma)}$. 
    \textbf{e)} Ensemble-averaged mean squared displacements (MSDs) over 6 to 9 trajectories showing superdiffusive behavior ($2 > \alpha > 1$) in the long time scale regime. White dashed lines are the linear fits between $3\langle\tau_\mathrm{R}\rangle$ to $11\langle\tau_\mathrm{R}\rangle$ with slope $\alpha$. Black solid lines denote ballistic ($\Delta t^2$) and diffusive ($\Delta t$) scaling. Inset: MSDs of individual particles for $\gamma=1.5$. The \textit{x}-axis is normalized by the mean run-time for each $\gamma$. Scale bars are 20.0  \textmu m in a-c, and 1.0\,\textmu m in the inset.}
    \label{fig:Pareto}
\end{figure}

We implemented these different power-law distributions in the run-times of ABPs while keeping the propulsion strength fixed, controlling only the propulsion direction via an external magnetic field programmed with an Arduino (see Methods). When particles are subjected to a uniform magnetic field, the hematite cube embedded in the TPM particle acquires a fixed orientation, maintaining a constant propulsion direction and resulting in straight runs under UV illumination, as can be seen in Fig.~\ref{fig:Pareto}a. A run ends when the rotation of the magnetic field reorients the particle. The run-times, or equivalently, the time between successive reorientations, are programmed to follow a predefined power-law distribution (see Equation \eqref{eq:pareto_dist}), while the particle’s new orientation is determined by the rotating magnetic field, which is designed to make all angles equally probable. By toggling the particle’s rotation between ON and OFF states, we guide the particles to follow run-lengths/run-times that conform to a power-law distribution.

Figure~\ref{fig:Pareto}a-c shows the overlaid experimental trajectories of ABPs executing Lévy walks for three different $\gamma$ values, clearly illustrating that the number of long runs decreases as $\gamma$ increases and vice versa. Supplementary Movie 1 further highlights the extent of spatial exploration over equal time durations for all three $\gamma$ values. The run-lengths were determined by first identifying the turning points (see Supplementary Note 3 for details regarding the detection of turning points) and then measuring the distance traveled by the particle between consecutive turns. We note that the mean particle speed was used to estimate run times from the measured run lengths.

Figure~\ref{fig:Pareto}d shows the normalized histograms of the run-time distributions, with a cutoff of $\tau_0 = 2$\,s. The log-log plot illustrates the power-law decay, revealing distinct differences across the three $\gamma$ values and highlighting the region of longer run-times. The solid lines represent the theoretical power-law curves for each $\gamma$ value. Due to long runs drawn from the power-law tail, particles often left the field of view, resulting in a limited number of recorded runs ($\sim 100$). In addition, small fluctuations in particle speed during each run further affected the run-time distribution. These factors lead to incomplete sampling of the power-law tail, particularly for $\gamma = 1.2$, and limit an exact match to the theoretical decay. However, the primary goal of this section is to demonstrate the tunability of run-lengths across different $\gamma$ values. This is further supported by the MSDs shown in Fig.~\ref{fig:Pareto}e. For all $\gamma$ values, at time scales $t \ll \langle \tau_\mathrm{R} \rangle$, the MSD exhibits ballistic behavior due to straight runs. At longer timescales, $t \gg \langle \tau_\mathrm{R} \rangle$, the MSD scaling indicates anomalous transport, aligning well with theoretical predictions and following the relation $\gamma + \alpha = 3$ \cite{zaburdaev2015levy}. The experimentally obtained long-time MSD scaling exponents were $\alpha = 1.6 \pm 0.1$, $1.4 \pm 0.1$, and $1.2 \pm 0.1$ for $\gamma = 1.2,\;1.5,$ and $1.8$, respectively, suggesting a quantitative agreement with theoretical expectations within a margin of error. For $\gamma = 1.2$, the deviation from the expected exponent is slightly larger and can be attributed to the difficulty of capturing the longest runs within the finite experimental field of view. In addition, we also analyze the velocity auto-correlation function (see Supplementary Note 5), which exhibits the expected scaling of $1-\gamma$ for $\gamma=1.5 \text{ and 1.8}$ but shows a larger deviation for $\gamma=1.2$ \cite{karani2019tuning}.

By tuning the parameter $\gamma$, we controlled the run-length distributions in simple ABPs, resulting in anomalous transport, as evident from the MSD scaling. Our ability to emulate diverse motility behaviors demonstrates that simple ABPs can be programmed to exhibit a broad spectrum of persistent random walks, significantly expanding beyond Lévy walks. Next, we encode three types of persistent random walks, namely RTP motion, Gaussian, and self-avoiding random walks. Lévy walks and RTP motion, both observed in \textit{E.\,coli}, exhibit striking similarities, with the key difference being that Lévy walks show superdiffusive behavior over longer timescales and display more extended straight runs.

\subsection*{Run-and-tumble, Gaussian and Self-avoiding walks}
\label{subsec:RTP_Gaussian_SAW}

\begin{figure}[H]
    \centering
    \includegraphics[width=1\linewidth]{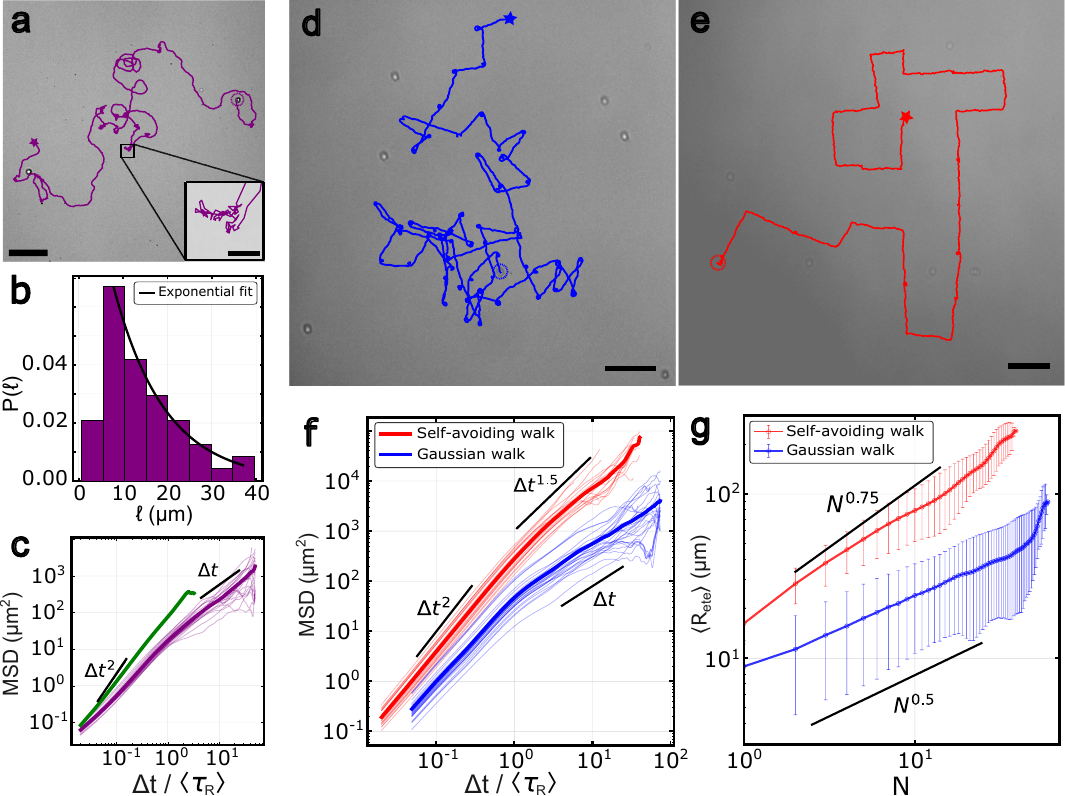}
    \caption{\textbf{Encoding three different persistent random walks.}
    \textbf{a-c) Run-and-Tumble motion.
    a)} Overlaid trajectory on a bright-field microscope image depicting the RTP motion of an Active Brownian Particle (ABP). Inset: Tumbling motion due to rotational diffusion.
    \textbf{b)} Normalized run-length distribution with the solid black curve representing the exponential fit. 
    \textbf{c)} Ensemble-averaged MSD of RTP over 19 trajectories exhibiting long-time diffusive scaling (purple) and MSD of only run segments of RTP exhibiting short-time nearly ballistic scaling (green). MSDs of the 19 individual particles are shown in the background. The \textit{x}-axis is normalized by the mean run-time.
    \textbf{d-g) Gaussian and Self-avoiding walk.
    d, e)} Overlaid trajectories of particles tracing a Gaussian walk (d) and a SAW (e). The star and circle denote the start and end points, respectively.
    \textbf{f)} Ensemble-averaged MSDs, approaching short-time ballistic and long-time diffusive scaling for Gaussian walks (averaged over 26 trajectories), while SAWs (averaged over 20 trajectories) exhibit short-time nearly ballistic and long-time superdiffusive behavior. The \textit{x}-axis is normalized by the mean run-time of the respective random walks. MSDs of individual particles are shown in the background.
    \textbf{g)} End-to-end distance vs. number of segments (runs), depicting a typical scaling of $N^{0.5}$ for Gaussian walks and a Flory scaling of $N^{0.75}$ for 2D SAWs. Error bars represent the standard deviation across 26 and 20 trajectories for Gaussian walk and SAW, respectively. Scale bars are 20.0\,\textmu m in a, d, and e, and 2.0\,\textmu m in the inset.}
    \label{fig:RTP_and_GvsS}
\end{figure}

Inspired by the RTP motion of \textit{E.\,coli}, where run-lengths typically follow an exponential distribution \cite{berg2004coli}, we implemented RTP-like behavior in spherical SPPs. To achieve this, we programmed the particle run-times (i.e., run-lengths) to follow a shifted exponential decay function \textit{P}($\tau_\mathrm{R}$) $\propto$ $e^{-(\tau_\mathrm{R}-\tau_\mathrm{c})/\tau_0}$ with a decay time $\tau_0=5$\,s and shift $\tau_\mathrm{c}=2$\,s, modeling the runs as ABPs in the absence of a magnetic field. The run-lengths were controlled by the duration for which the UV light remained ON, while tumbles were induced by turning the UV light OFF for a fixed period longer than the rotational relaxation time, allowing the particle to reorient purely through Brownian rotational diffusion via thermal fluctuations. The particle trajectory was entirely controlled by toggling the UV light, without requiring a magnetic field for reorientation (see Methods).

The overlaid trajectory of the RTP motion is shown in Fig.~\ref{fig:RTP_and_GvsS}a, where the tumble motion is evident from the dark “knots” along the non-straight runs driven by the particle’s rotational diffusion. Supplementary Movie 2 shows RTP motion executed by multiple particles simultaneously. A magnified view of the tumbling, shown in the inset of Fig.~\ref{fig:RTP_and_GvsS}a, shows a clear picture of Brownian motion. For {\it {E. coli}}, the typical timescale of runs ($\sim$1\,s) is an order of magnitude larger than that of tumbles ($\sim$0.1\,s) \cite{berg2004coli}. However, in our model, since tumbling is purely governed by rotational diffusion, the reorientation must persist longer than the rotational relaxation time. Consequently, we set the tumble time to 10\,s, exceeding the rotational relaxation time of around 6\,s. Regardless of the absolute run-lengths, the exponentially distributed run-lengths (see Supplementary Note 3 for details regarding the detection of turning/tumbling points), as shown in Fig.~\ref{fig:RTP_and_GvsS}b, lead to long-time diffusive scaling, as depicted in the ensemble-averaged MSD in Fig.~\ref{fig:RTP_and_GvsS}c. 
The short-time MSD, however, exhibits sub-ballistic scaling because the tumble time is slightly longer than the average run time in our implementation. When only the run segments are considered, the MSD shows nearly ballistic scaling, consistent with theoretical expectations for RTP motion (green line in Fig.~\ref{fig:RTP_and_GvsS}c).

Random walks not only describe the motion of many living organisms, but are also extensively used to study the structural properties of polymers, optimizing network exploration, analyzing financial markets, and many more \cite{costa2007exploring, flory1953principles, fama1995random}. Consequently, we focus on two models commonly used in polymer physics, namely, Gaussian chains and SAWs. While the Gaussian chain provides a basic framework based on simple random walks, the SAW model offers a more realistic description by incorporating volume exclusion between monomers.

In a Gaussian polymer chain model, monomer-monomer bond lengths follow a Gaussian distribution. To represent this as a random walk, we map bond lengths to run-lengths and bond angles to turns. Since the particle speed remains constant, the run-time and run-length distributions have the same form. The Gaussian run-time distribution is given by

\begin{equation}
    P(\tau_\mathrm{R}) = \frac{1}{\sqrt{2\pi}\sigma}\exp\left[-\frac{(\tau_\mathrm{R}-\mu)^2}{2\sigma^2}\right] \,\text{,}
    \label{eq:gaussian_distribution}
\end{equation}

\noindent
where $\sigma^2$ is the variance and $\mu=\langle \tau_\mathrm{R} \rangle$ is the mean run-time. The orientation of each step can take any possible value irrespective of the previous steps, making the run-lengths uncorrelated according to a Markovian model \cite{billingsley1961statistical}. The mean end-to-end distance of a Gaussian chain scales with the number of steps as $\langle R_\mathrm{ete} \rangle \propto N^{0.5}$, leading to diffusive scaling of the MSD at long time scales.
In contrast, the SAW model is characterized by a non-Markovian trajectory, where each step is constrained to avoid overlap with previous steps. This self-avoiding behavior yields a random walk that is spatially more spread out, resulting in an end-to-end distance that scales with the number of steps as $\langle R_\mathrm{ete} \rangle \propto N^{\nu}$, where $\nu=3/(2+d)$ is the Flory exponent, and \textit{d} is the dimension of the system \cite{flory1953principles}. For a two dimensional system, $\langle R_\mathrm{ete} \rangle \propto N^{3/4}$.

To implement Gaussian walk-like behavior in ABPs, the run-times were randomly drawn from a Gaussian distribution (Equation \eqref{eq:gaussian_distribution}) with $\mu = 2$\,s, $\sigma = 0.5$\,s and a lower bound at 0, such that $P(\tau_\mathrm{R}<0)=0$. The straight runs were driven by ABPs under a stationary magnetic field, while turns were controlled by a rotating magnetic field according to angles sampled from a uniform distribution (see Methods). Figure~\ref{fig:RTP_and_GvsS}d depicts the overlaid trajectory of an ABP undergoing a Gaussian walk, and  Supplementary Movie 3 demonstrates several particles exhibiting Gaussian walks.

To achieve SAWs, the motion was restricted to a square lattice, where turn angles were limited to \{-90\degree, 0\degree, 90\degree\} to reproduce the characteristic statistics of SAWs with a minimal model. In this case, run-times were fixed at $\tau_\mathrm{R} = 5$\,s to maintain equal run-lengths (see Supplementary Note 3 for details regarding the detection of turning points). Figure~\ref{fig:RTP_and_GvsS}e illustrates a typical experimental trajectory of a single particle exhibiting SAW behavior; Supplementary Movie 4 shows multiple particles simultaneously displaying SAWs. This further highlights how the encoded random walks evolve independently, in contrast to purely magnetic-field-driven systems, where all particles follow the exact same trajectory path. This inherent stochasticity, arising from the initial self-propulsion direction of individual particles and their reorientation dynamics, introduces randomness into the system. 

A key difference between the two walks is the presence of crosswalks in the Gaussian walk trajectory as shown in Fig.~\ref{fig:RTP_and_GvsS}d and \ref{fig:RTP_and_GvsS}e. The ensemble-averaged MSDs for Gaussian walks and SAWs display diffusive and superdiffusive scaling, respectively, at longer times, with the MSDs of multiple particle trajectories following a similar scaling, as shown in Fig.~\ref{fig:RTP_and_GvsS}f. In the short-time regime, MSDs of both Gaussian walks and SAWs exhibit nearly ballistic scaling as they have frequent runs. The end-to-end distance as a function of the number of steps, as shown in Fig.~\ref{fig:RTP_and_GvsS}g, shows excellent agreement with theory, with $N^{0.5}$ scaling for the Gaussian walk and $N^{0.75}$ scaling for the SAW, obeying the Flory scaling in two dimensions \cite{flory1953principles}.

Figure~\ref{fig:RTP_and_GvsS} demonstrates the successful implementation of Gaussian walks and SAWs in ABPs, with the end-to-end distance as a function of the number of steps showing good agreement with theoretical predictions.

\subsection*{Dynamic switching between random walk modes}
\label{subsec:switchability}

\begin{figure}[ht]
    \centering
    \includegraphics[width=1\linewidth]{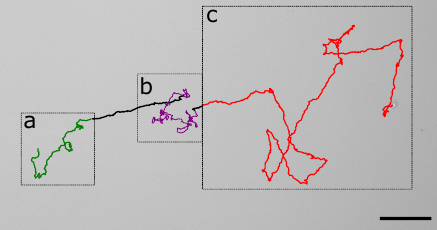}
    \caption{\textbf{Switching between three different types of random walks. a)} RTP motion, \textbf{b)} Gaussian walk, and \textbf{c)} L\'evy walk, demonstrating the flexibility by switching between different modes. The black trajectories between successive walks act as a transition. RTP motion and Gaussian walk are separated by a straight run using a stationary magnetic field, while Gaussian and L\'evy walks are separated by Brownian diffusion. The RTP motion and Gaussian walks correspond to 60\,s runs and the L\'evy walk was run for 180\,s in order to show longer run-lengths. Scale bar is 20.0\,\textmu m.}
    \label{fig:switchability}
\end{figure}

Figures \ref{fig:Pareto} and \ref{fig:RTP_and_GvsS} illustrate the precise control over particle motion achieved by encoding different random walks into ABPs. We further demonstrate the versatility of our system by incorporating multiple types of walks on-demand within a single particle, without the need to modify the setup. The RTP motion and Gaussian walks were each conducted for 60\,s, while the Lévy walk was run for approximately three times longer to emphasize its extended run-lengths, as shown in Fig.~\ref{fig:switchability} (see Supplementary Movie 5). All operations were pre-programmed and executed using an Arduino microcontroller.
This ability to dynamically switch between complex propulsion patterns paves the way for potential applications in fields such as targeted drug delivery and network exploration.

\subsection*{Light-modulated potential landscapes}
\label{subsec:potential_landscapes}
\begin{figure}[ht]
    \centering
    \includegraphics[width=0.84\linewidth]{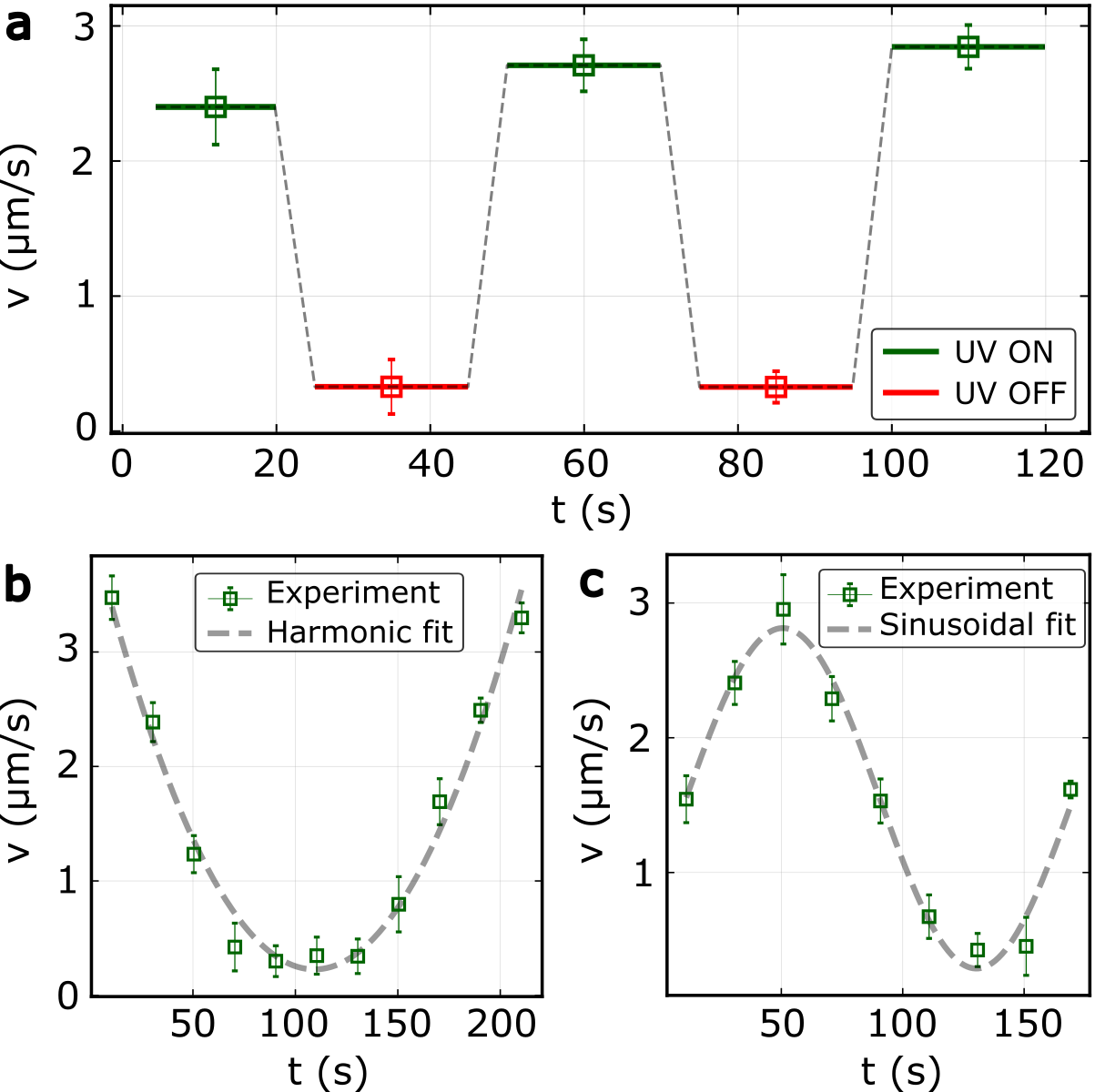}
  \caption{\textbf{Generating local velocity landscapes. a)} Speed vs. time plot showing active (ON) and passive (OFF) states, where each state lasts for 25\,s. The gray dashed line represents the transition process. \textbf{b, c)} Speed vs. time plot of a particle experiencing a locally generated harmonic and sinusoidal velocity landscape. The gray dashed lines represent the curve fits to respective functions. The plots present 20\,s segments after applying a 5\,s moving average, with error bars indicating the standard deviation within each segment.}
    \label{fig:potential_landcapes}
\end{figure}
We demonstrate how to modulate the velocity of particles without requiring complex setups like optical tweezers \cite{juan2011plasmon,ghosh2019all}. To achieve this, we first fix the orientation of the particle by applying a constant stationary magnetic field ($\sim$ 2\,mT), ensuring that the particle moves along a straight line. This allows us to generate complex velocity landscapes by simply varying the intensity of the UV light to control particle speed. Unlike optical tweezers, which generate a spatially fixed potential landscape within the sample cell, our approach allows each particle to experience a velocity landscape within its own frame of reference, regardless of its location.

Under conditions where rotational diffusion was negligible, we assessed variations in propulsion speed along a straight-line trajectory. Figure~\ref{fig:potential_landcapes}a shows the particle’s response to toggling the UV light ON and OFF every 25\,s. When the UV state changes, the particle speed adapts to the new light intensity and remains stable until the next toggle. The mean and standard deviation correspond to a 5\,s moving average of the particle speeds. Additionally, using this approach, we created harmonic and sinusoidal trends in the velocity, enabling particles to move through these landscapes as shown in Fig.~\ref{fig:potential_landcapes}b and \ref{fig:potential_landcapes}c. The propulsion speed of the particle accurately reproduces the input parameters for these landscapes, specifically reaching the minima of the harmonic curve (within 0.6\%) and matching the time period of the sinusoidal curve (within 8.8\%).

\subsection*{Steering ABPs into complex paths}\label{subsec:shapes}

\begin{figure}[h]
    \centering
    \includegraphics[width=1\linewidth]{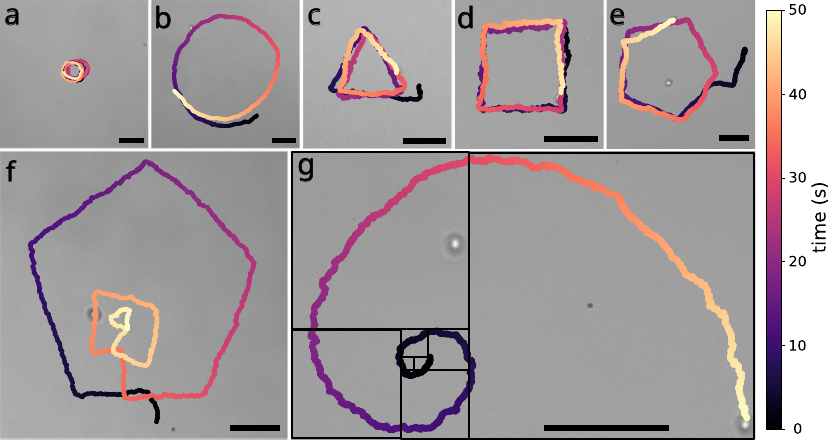}
    \caption{\textbf{Steering particles along complex geometric paths. a, b)} Circles with high and low rotational frequencies, respectively. \textbf{c)} Triangle, \textbf{d)} Square, \textbf{e)} Pentagon, \textbf{f)} Nested polygons, and \textbf{g)} Fibonacci spiral. The color bar shows the time in seconds. Scale bars are 10.0\,\textmu m.}
    \label{fig:shapes}
\end{figure}

Finally, by leveraging precise control over the propulsion direction of ABPs, we steer them to follow shape patterns ranging from simple to complex, as shown in Fig.~\ref{fig:shapes}. When we apply a uniformly rotating external magnetic field, our ABPs become chiral active particles, exhibiting persistent circular trajectories. To quantify this motion, we measured both the mean squared displacement (MSD) and the mean squared angular displacement (MSAD). These circular swimmers display a characteristic oscillatory signature in their MSDs \cite{vutukuriRationalDesignDynamics2017}, which distinguishes them from the other propulsion modes discussed above (see Supplementary Note 7). In contrast, the MSAD increases continuously over time, consistent with persistent rotational motion. By adjusting the rotational frequency of the rotating magnetic field, we controlled the radius of their circular trajectories, achieving a range of sizes and levels of compactness (Fig.~\ref{fig:shapes}a,b). Additionally, by applying discrete rotations of the magnetic field at periodic intervals, we guided the particles into well-defined polygonal shapes such as triangles, squares, and pentagons, as shown in Fig.~\ref{fig:shapes}c-e. Building on this approach, we further steered particles through intricate patterns, including nested polygons and spirals inspired by the Fibonacci series \cite{cameron1994combinatorics}, as depicted in Fig.~\ref{fig:shapes}f,g. To showcase the versatility of our method, we implemented real-time maneuvering of particle trajectories using joystick control (see Supplementary Note 6), demonstrating accurate navigation along predefined paths. Additional details on the implementation of each shape can be found in the Methods section.

Supplementary Movie 6 illustrates all the shapes and patterns used to steer ABPs. This level of directional control further demonstrates the adaptability of our system to generate complex and programmable motion patterns.

\subsection*{Clustering dynamics of ABPs and circular swimmers}\label{subsec:collective}

\begin{figure}[hbtp]
\centering
\includegraphics[width=1\linewidth]{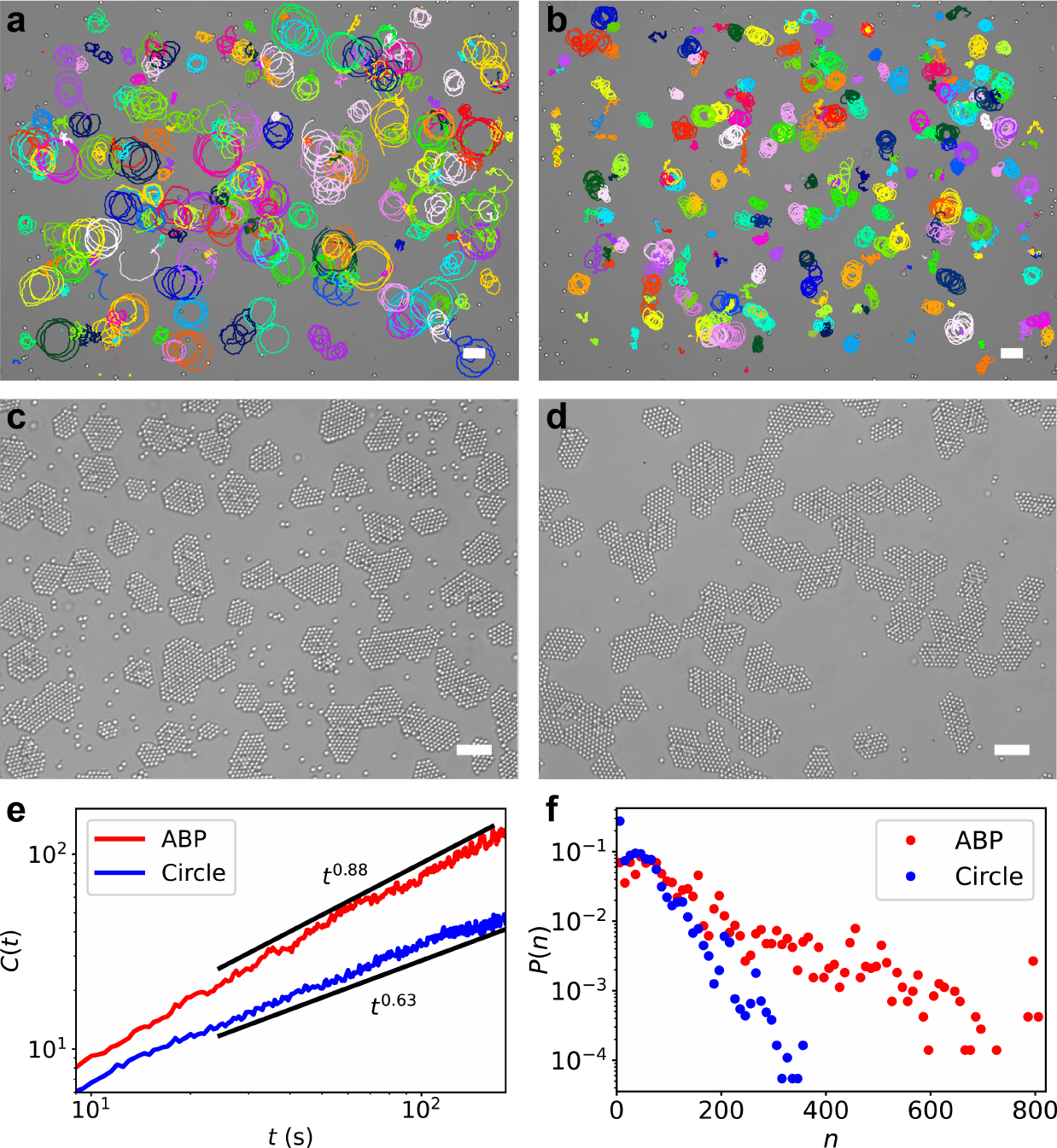}
\caption{\textbf{Clustering dynamics of circular swimmers and ABPs.} \textbf{a, b)} Overlaid trajectories on bright field microscope images revealing the particles moving in circles with diameters \( D_1 = \) $12.2 \pm 3$ \textmu m and \( D_2 = \) $5.8 \pm 1.3$ \textmu m, respectively. Tracks are shown over a 20-second period, with different colors denoting individual particle trajectories. \textbf{c, d)} At high area fraction (\( \phi = 0.26 \)), circular swimmers and ABPs self-organize into a hexagonal lattice. \textbf{e)} Time evolution of the average cluster size \( C(t) \) reveals sublinear growth for circular swimmers, following \( C(t) \sim t^{\beta} \) with \( \beta \approx 0.63 \), while ABPs exhibit faster growth with \( \beta \approx 0.88 \). Solid black lines indicate the fitted scaling trends. \textbf{f)} Cluster size distributions measured after the same time duration ($t=3$ min) show that ABPs form substantially larger crystalline domains than circular swimmers. Scale bars are 10.0\,\textmu m.}
\label{fig:Collective dynamics}
\end{figure}

To demonstrate the versatility of our system, we focus on the clustering dynamics of chiral active particles and compare them with ABPs. We used 1 \textmu m TPM particles embedded with 0.4 \textmu m hematite cubes, which suppress induced magnetic dipole interactions between particles (see Methods), ensuring that the clustering behavior is driven solely by particle motion and its associated hydrodynamic and phoretic interactions. Motivated by prior simulation studies predicting hyperuniformity in systems of circular microswimmers \cite{huang2021circular,torquato2021swimming}, we first demonstrate the tunability of the circular path diameter by varying the rotating magnetic field frequency from 0.2 Hz to 0.44 Hz. This results in circular trajectories with diameters of $12.2 \pm 3$ \textmu m and $5.8 \pm 1.3$ \textmu m, respectively. At both frequencies, particles followed stable circular paths with negligible interparticle interactions, as shown in the overlaid trajectories in Fig.~\ref{fig:Collective dynamics}a,b.

We used the circular swimmers with $5.8 \pm 1.3$  \textmu m diameter trajectories to probe clustering behavior. To compare their clustering dynamics with those of ABPs, we fixed the area fraction at $\phi = 0.26$ and studied both systems under identical conditions. Time-resolved observations revealed that both systems initially formed small clusters, which gradually grew into larger, hexagonally packed crystalline domains (Fig.~\ref{fig:Collective dynamics}c,d). However, clusters formed by circular swimmers grew more slowly compared to ABPs, as evident in the average cluster size (see Supplementary Note 8). We observe a sublinear growth law in the average cluster size: $C(t) \sim t^\beta$, with $\beta = 0.63$ for circular swimmers, compared to $\beta = 0.88$ for ABPs (Fig.~\ref{fig:Collective dynamics}e). Correspondingly, the cluster size distribution confirmed that ABPs formed significantly larger crystalline domains than circular swimmers after approximately 3 minutes (Fig.~\ref{fig:Collective dynamics}f). Further analysis of the number of particles in the largest cluster reveals that while ABP clusters continued to grow, the largest cluster formed by circular swimmers did not grow beyond a critical size (see Supplementary Fig. S9).

\section*{Conclusions}
In this work, we developed an experimental platform for encoding various persistent random walks into self-propelled particles (SPPs), significantly expanding the accessible range of motion behaviors beyond conventional active Brownian particles (ABPs). By controlling the interplay between propulsion direction via an external magnetic field and propulsion strength through light intensity, we successfully implemented diverse motion patterns, starting with Lévy walks featuring three different power-law run-length distributions. The mean-squared displacements (MSDs) exhibited anomalous diffusion, $\langle r^2 \rangle \propto t^\alpha$, where the exponent $\alpha$ was strongly influenced by the run-length power-law exponent $\gamma$. For all three $\gamma$ values, the experimentally measured $\alpha$ values showed good agreement with theoretical predictions, with the highest degree of superdiffusion observed at $\gamma = 1.2$ \cite{zaburdaev2015levy}.

For run-and-tumble (RTP) motion and Gaussian walks, the long-time MSD scaling followed the expected diffusive behavior, whereas for self-avoiding walks (SAWs), we observed superdiffusive scaling. Additionally, our system enabled measurements of end-to-end distance versus the number of steps, yielding scaling laws that quantitatively matched predictions for Gaussian chains and reproduced the Flory scaling for SAWs \cite{flory1953principles}, further reinforcing the robustness of our approach. The platform also allows rapid, on-demand switching between different motion modes within a single experiment, offering unprecedented flexibility for studying active particle dynamics. Moreover, we demonstrated how spatiotemporal potential landscapes can be used to guide particle trajectories.
Beyond controlling particle motion, we demonstrated the ability to steer particles along well-defined paths, ranging from simple circles and polygons to complex structures, including nested polygons and Fibonacci spirals. Unlike purely magnetic-field-steered systems, where all particles follow a common trajectory \cite{shen2023magnetically,estelrich2015iron}, the initial propulsion direction of individual ABPs and fluctuations in propulsion strength arising from the chemical reaction introduce stochasticity into our system (see Supplementary Note 4), allowing each encoded random walk to evolve independently. This inherent stochasticity makes our approach fundamentally distinct and biologically relevant, capturing the diversity of natural motility patterns observed in microorganisms. Moreover, due to this stochasticity, our particles may be effectively described by the active Ornstein–Uhlenbeck particle (AOUP) model, in which the dynamics follow a Langevin equation with a propulsion force governed by an Ornstein–Uhlenbeck process \cite{kimActiveDiffusionSelfPropelled2024,hanNonequilibriumDiffusionActive2023}.

Finally, going beyond single-particle trajectories, we demonstrated that differences in propulsion modes can influence the clustering behavior of active colloids, as demonstrated through the comparison between ABPs and circular swimmers. In our experiments, circular swimmers exhibited slower cluster growth than ABPs, providing an example of how encoded particle-level dynamics can shape emergent behavior. At dilute concentrations, circular swimmers followed stable trajectories with negligible interparticle interactions, suggesting that suppressing phoretic clustering in future designs could establish them as ideal experimental models for probing the hyperuniform states predicted in recent simulations and observed in biological microswimmer assemblies \cite{huang2021circular,torquato2021swimming}.

More broadly, our findings establish a versatile framework for investigating complex transport phenomena, adaptive search strategies, and dynamic self-organization. Future studies could extend this approach to explore collective behaviors arising from interactions between particles programmed with different types of random walks, offering insights into the emergent dynamics of both biological and synthetic active systems.

\section*{Methods}\label{sec:Methods_and_materials}

\subsection*{Synthesis of hematite cubes}\label{subsec:particlesynthesis}
Hematite cubes were synthesized following previously reported methods \cite{sugimoto1992preparation, shelke2023exploiting}. To prepare 0.9\,\textmu m cubes (see SEM image in Supplementary Note 1), a 50\,mL solution of 2\,M FeCl$_3$·6H$_2$O (ACS, 97.0–102.0\%, Sigma-Aldrich) was prepared in a 250\,mL Pyrex bottle. While vigorously shaking the bottle, 50\,mL of 5\,M NaOH was added over 20\,s, followed by continuous stirring for 10\,min. The mixture was then transferred to a preheated oven at 100\,\textdegree C and maintained at this temperature for 8 days. To remove any unreacted chemicals, the resulting cubes were washed multiple times by centrifugation and redispersion, first in Milli-Q water (18.2 M$\Omega \cdot$cm, Milli-Q\textsuperscript{\textregistered} EQ 7000), then in ethanol, and finally again in Milli-Q water. To synthesize 0.4 \textmu m cubes for the clustering experiments, the same protocol was followed, except that NaOH was added to the reaction mixture over a period of 5 s.

\subsection*{Synthesis of photocatalytic microswimmers}\label{subsec:TPMparticlesynthesis} 
Photocatalytic microswimmers, composed of a hematite cube partially embedded in a spherical TPM particle, were synthesized following a modified version of established protocols \cite{aubret2021metamachines, sacanna2013shaping}. A 1\,mL solution of $\sim$0.05\%\,(w/v) hematite particles was added to 100\,mL of Milli-Q water in a glass beaker and sonicated for 5\,min, resulting in a reddish color. Under mild stirring (100\,rpm), 240\,\textmu L of 28\% NH\textsubscript{4}OH (28–30\%, Sigma-Aldrich) was added to the beaker, followed by 250\,\textmu L of 3-(Trimethoxysilyl)propyl methacrylate (TPM, Sigma-Aldrich). The beaker was covered with Parafilm\textsuperscript{\textregistered} and stirred for 30\,min, resulting in a pinkish solution that indicated the formation of hematite cube-embedded TPM particles. The particle size was tuned by sequentially adding 50\,µL of TPM every 15\,min under continuous stirring, while monitoring the size with a bright-field microscope. Once the target size (1.5\,\textmu m, see Supplementary Note 1) was achieved, the particles were stabilized by adding 0.5\,mL of 5\,wt\% Pluronic F-108 solution, followed by 2\,min of stirring.

At this stage, the stabilized particles had the hematite cube almost entirely encapsulated by the TPM droplet. To partially expose the cube, dewetting from the TPM phase was induced by gradually lowering the pH through incremental additions of 0.5\,mL of 1\,M HCl (Sigma-Aldrich), followed by 3\,min of stirring, while monitoring the process with a bright-field microscope. Once the desired degree of dewetting was achieved, 50\,mL of 0.2\,mg mL\textsuperscript{-1} azobisisobutyronitrile (AIBN, Sigma-Aldrich) solution was added and stirred for 2\,min to initiate polymerization. The mixture was then transferred to a 500\,mL Pyrex bottle, diluted to 400\,mL with Milli-Q water, and heated in an 80\,\textdegree C oven for 3\,h. The particles were subsequently sedimented under gravity overnight, after which the supernatant was removed. Finally, they were washed and purified via centrifugation to remove secondary nucleated particles and unreacted chemicals. To synthesize 1 \textmu m TPM particles embedded with 0.4 \textmu m hematite cubes, the same protocol was used, except with 0.4 \textmu m cubes, and TPM growth was stopped at 1 \textmu m based on optical microscopy measurements.

\subsection*{Sample preparation and microscopy imaging}\label{subsec:Sample preparation}
The particles were imaged using custom-made observation chambers assembled on 24\,x\,50\,mm \#1.5 cover glasses (VWR). Before assembly, the coverslips were plasma cleaned for 20\,min using a plasma cleaner (Bioforce Nanosciences, UV/Ozone ProCleaner Plus). To construct the chamber, the wide end of a 200\,\textmu L pipette tip was cut off and glued onto the coverslip with a UV-curable adhesive (Norland Optical Adhesive 68), followed by UV curing for 5\,min. 

A 30\,\textmu L dilute aqueous suspension of particles ($\sim$100 particles per \textmu L) containing 5\,mM sodium hydroxide (NaOH, Fisher Chemical) and 6\%\,v/v hydrogen peroxide (H\textsubscript{2}O\textsubscript{2}, Fisher Scientific) was prepared, added to the chamber, and allowed to sediment for 2–3 minutes. The sample was then equilibrated for 2 minutes before the experiment was initiated. The chamber was then sealed with Parafilm\textsuperscript{\textregistered}, placed on the microscope stage, and imaged using a Nikon Eclipse TE2000-U inverted fluorescence microscope equipped with a 40× objective (N.A. 0.75) and a Basler acA4112–30um CMOS camera. Time-lapse image sequences were recorded in bright-field mode at 10 frames per second (fps). Simultaneously, the sample was illuminated through the objective lens with ultraviolet (UV) light using a 365\,nm LED (Thorlabs Solis\textsuperscript{\textregistered} SOLIS-365C High-Power LED), with an adjustable intensity ranging from 0 to 30\,mW cm\textsuperscript{-2}.

\subsection*{Control of particle motion and orientation}\label{subsec:heterodimer control}
The motion of the particles was controlled by independently tuning their propulsion speed and orientation (see Supplementary Note 2 for a schematic depiction of the setup). For hematite cube-embedded TPM particles, the protruding hematite cube serves as a photocatalyst, decomposing hydrogen peroxide upon UV illumination. This reaction generates a local chemical concentration gradient of constituent molecules, driving self-diffusiophoretic propulsion \cite{palacci18}.

The particles were propelled along their symmetry axis, with motion directed toward the hematite cube. Their speed was modulated by varying the UV light intensity, which was controlled via a UV LED driver (Solis\textsuperscript{\textregistered} LED Driver) using an external transistor-transistor-logic (TTL) signal. By adjusting the duty cycle of the TTL signal via an Arduino, the current through the LED driver was regulated, thereby tuning the UV intensity. The UV light could be set to a constant level or varied dynamically over time. At zero UV intensity, the particles exhibited purely Brownian motion, while at maximum intensity, they reached a propulsion speed of up to 4.5\,\textmu m s\textsuperscript{-1} in 6\%\,v/v H\textsubscript{2}O\textsubscript{2} solution. Although a fixed UV light intensity was maintained in the experiments to ensure a constant propulsion speed, the particles exhibited a speed distribution fluctuating around a mean speed $\langle v \rangle$ (see Supplementary Note 4). This variation may stem from the polydispersity of the hematite cube and the surface heterogeneity of the substrate.

The orientation of the particles was controlled via the weak intrinsic magnetic moment of the hematite cube \cite{meijer2021}, which aligned with an externally applied magnetic field.
To generate and control this field, a permanent magnet was mounted on the rotor of a servo motor (Parallel Feedback 360\textdegree\,High-Speed Servo), positioned 5\,cm above the sample. The magnet's north-south axis was oriented parallel to the substrate and perpendicular to the motor's rotation axis, which was centered over the sample using a screw gauge to minimize drift. When the magnetic field was rotated, the cube followed synchronously (with a field strength of approximately 1–2\,mT in the sample plane), enabling precise control over its orientation. The rotation speed and direction of the magnetic field were controlled using pulse width modulation (PWM). The motor allowed for tunable angular velocities in both clockwise and counterclockwise directions, with rotation speeds of 0.2-3\,Hz. An Arduino Uno R3 microcontroller with a PWM shield (Adafruit 16-Channel PWM/Servo Shield) was used to provide the TTL signal and the PWM signal, which controls the propulsion strength and direction, respectively. 

To investigate different modes of particle motion, we implemented various trajectory algorithms governing the magnetic field orientation and the UV light intensity. In each case, the motion consisted of alternating straight runs and turns, with specific rules governing the duration and nature of each phase. The applied magnetic field was either held stationary, dynamically adjusted, or entirely absent, depending on the mode of random walk implemented in the experiment. The UV light intensity was kept constant in all experiments except for run-and-tumble particle (RTP) motion, where it was cycled on and off according to predefined time distributions. Below, we provide additional details about the different motion strategies employed for: L\'evy walks, RTP motion, Gaussian walks, Self-avoiding walks (SAWs), and Shapes.

\begin{itemize}
    \item \textbf{L\'evy walk}: L\'evy walks consisted of turns lasting between 0.25\,s and 0.75\,s, and were drawn from a uniform distribution. The magnet rotated at approximately 2\,Hz with a fixed direction, resulting in turn angles distributed between $\pi$\,and 3$\pi$.

    \item \textbf{RTP}: For RTP motion, the run-times, corresponding to when the UV light was on, were drawn from a shifted exponential decay function, which ensured a low probability of very short run-lengths, which would have resulted in noisier trajectories. This resulted in run times between 0.1 and 7.9 s.

    \item \textbf{Gaussian walks}: Similar to L\'evy walks, Gaussian walks also consisted of alternating runs and turns governed by the magnet orientation. The duration of turns was slightly longer, chosen from a uniform distribution in the range of 0.5\,s to 1.0\,s, with a constant rotation speed of 2\,Hz.

    \item \textbf{SAWs}: SAWs were implemented using a series of straight runs connected by turns at fixed angles. The choice of turn was constrained by the particle’s trajectory history, ensuring adherence to self-avoidance rules derived from a 2D lattice simulation of SAWs. This digital implementation of volume exclusion in the path simulation ensured that the particle did not overlap with its previously traveled path.

    \item \textbf{Circles}: The circle diameter was inversely proportional to the magnet’s rotation frequency, with higher frequencies producing smaller circles and lower frequencies resulting in larger ones.
    
    \item \textbf{Polygons}: The particles were steered through a sequence of straight runs and sharp turns, each with a fixed run-time and turn angle. Triangles, squares, and pentagons were obtained by setting the turn angles to 120\textdegree, 90\textdegree, and 72\textdegree, respectively.

    \item \textbf{Fibonacci Spirals}: The particles were steered through a series of quarter-circle arcs with radii following the Fibonacci sequence. The radius of each arc was increased by decreasing the magnet’s rotation frequency while adjusting the turn-time accordingly. To maintain proper frequency ratios, an effective duty cycle was applied to the servo.

    Each quarter turn consisted of nine periods, with each period including 125\,ms of rotation followed by a delay, in which the magnet orientation was stationary. The rotation speed was kept constant at 4.5\,s per full rotation, while the delay time increased progressively to reduce the effective rotation frequency:
    \begin{itemize}
        \item First two quarter-circle arcs: 9 periods of 125\,ms rotation with 0\,ms delay (\(9 \times 125 = 1125\) ms, forming a quarter turn at 4.5\,s per rotation).
        \item Third quarter-circle arc: 9 periods of 125\,ms rotation, each followed by a 125\,ms delay, doubling the effective radius.
        \item Subsequent quarter-circle arcs: The delay was increased to 250\,ms for the fourth quadrant, 500\,ms for the fifth, and so on. Here, the rotation time (125\,ms) plus the delay time of each quadrant follows the Fibonacci series (125\,ms, 125\,ms, 250\,ms, 375\,ms, 625\,ms, ...).
    \end{itemize}
\end{itemize}

\subsection*{Particle tracking and analysis}
The particles were tracked from bright-field microscopy movies using the Difference of Gaussians (DoG) detector in the TrackMate plugin in Fiji (ImageJ) \cite{schindelin2012fiji,tinevez2017trackmate}. The detected particle positions were then analyzed to determine run-lengths by identifying turning points (see Supplementary Note 3). All quantitative analyses, including run-time and run-length distributions, MSDs, end-to-end distributions, velocities, overlaid trajectories, and curve fitting, were performed using custom Python scripts incorporating existing library packages.

\section*{Data availability}

The data supporting the findings of the manuscript are available in the 4TU.ResearchData repository at https://doi.org/10.4121/98fc75ad-ca7a-4d3b-adee-4f32285a4fcd.

\section*{Code availability}
The analysis codes used in this study are available in the 4TU.ResearchData repository at https://doi.org/10.4121/98fc75ad-ca7a-4d3b-adee-4f32285a4fcd.

\section*{Acknowledgements}
The authors thank M. Claessens for kindly providing access to a fluorescence microscope and I. Punt for her help with the SEM measurements. Shravya Narendula for useful discussions. H. R. Vutukuri acknowledges funding from The Netherlands Organization for Scientific Research (NWO-M1-OCENW.M.21.309, \&
NWO-OCENW.XS23.4.115).

\section*{Author contributions}
H.R.V. conceived and supervised the project, and secured funding. All authors contributed to project design. Y.S. synthesized the particles, and S.v.d.H. set up the magnetic field. T.S., Y.S., S.v.d.H., and A.N.S. performed the experiments. H.R.V. and T.S. wrote the manuscript. All authors participated in discussions, data analysis, and reviewed and edited the manuscript.

\section*{Competing interests}
The authors declare no competing interests.

\clearpage
\newpage

\begin{center}
\textbf{\large Supplementary Information}
\end{center}

% Reset counters
\setcounter{equation}{0}
\setcounter{figure}{0}
\setcounter{table}{0}
\setcounter{page}{1}
\setcounter{section}{0}

% Add 'S' prefix to equations, figures, tables, and sections
\makeatletter
\renewcommand{\theequation}{S\arabic{equation}}
\renewcommand{\thefigure}{S\arabic{figure}}
\renewcommand{\thetable}{S\arabic{table}}
\renewcommand{\thesection}{S\arabic{section}}
\makeatother

%%%%%%%%%%%%%%%%%%%%%%%%%%%%%%%%%%%%%%%%%%%%%%%%%%%%%%%%%%%%%%%%%%%%%%%%%%%%%%%%%%%%%%%%%%%%%%%%%%%
%-------------------------------------SUPPLEMENTARY INFORMATION-----------------------------------%
%%%%%%%%%%%%%%%%%%%%%%%%%%%%%%%%%%%%%%%%%%%%%%%%%%%%%%%%%%%%%%%%%%%%%%%%%%%%%%%%%%%%%%%%%%%%%%%%%%%
%\setcounter{section}{0}
%\renewcommand{\thesection}{S\arabic{section}}
%\renewcommand{\thefigure}{S\arabic{figure}}
%\renewcommand{\theequation}{S\arabic{equation}}
%\title[Article Title]{\centering{\bf Supplementary Information:\\ \bf{Programmable Persistent Random Walks in Active Brownian Particles Govern Emergent Dynamics}}}

%\author[1]{\fnm{Tarun} \sur{Sunkesula Raghavendra}}
%\author[1]{\fnm{Yogesh} \sur{Shelke}}
%\author[1]{\fnm{Stijn} \sur{van der Ham}}
%\author[1]{\fnm{Anpuj} \sur{Nair S}}
%\author*[1]{\fnm{Hanumantha Rao} \sur{Vutukuri}}\email{h.r.vutukuri@utwente.nl}
%\affil[1]{\orgdiv{Active Soft Matter and Bio-inspired Materials Lab, Faculty of Science and Technology, MESA+ Institute for Nanotechnology and Center for Brain-Inspired Nano Systems}, \orgname{University of Twente}, \orgaddress{\city{Enschede}, \postcode{7500 AE}, \country{The Netherlands}}}

\maketitle

\label{Supplementary}

\section*{Supplementary Note 1: SEM characterization of particles}\label{SM:SEM}
The morphology of the particles was characterized using scanning electron microscopy (SEM). To prepare SEM samples, a solution of particles in Milli-Q water was dried onto a silicon wafer, which was then coated with a 5\,nm platinum/palladium layer. Imaging was performed using a JEOL JSM-6010A SEM at 5 kV.

\begin{figure}[H]
    \centering
    \includegraphics[width=0.75\linewidth]{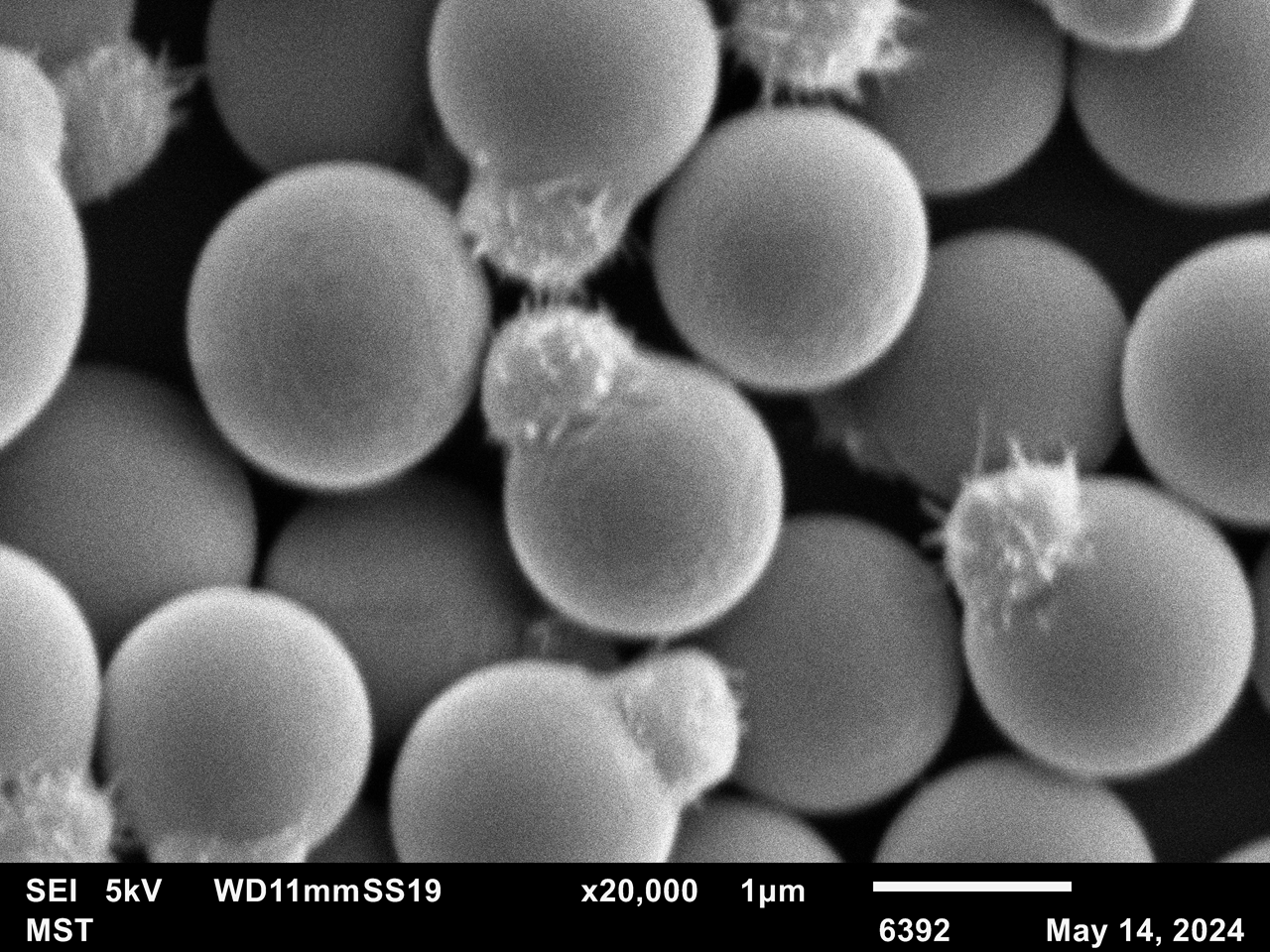}
    \caption{Scanning electron microscope image of the hematite cube-embedded TPM particles.}
    \label{fig:SEMimage}
\end{figure}

As shown in Fig.~\ref{fig:SEMimage}, SEM images at 20,000× magnification reveal particles consisting of a cubic hematite head partially embedded in a spherical TPM body, with most of the cube’s surface area exposed ($\sim$20-25\% embedded). The average diameter of the TPM body was measured to be 1.5\,$\pm$\,0.1\,\textmu m, based on the measurement of 100 particles using ImageJ FIJI \cite{schindelin2012fiji}.

\clearpage

\section*{Supplementary Note 2: Experimental setup}

Particle dynamics were imaged in a custom-made observation chamber using bright-field microscopy. The orientation of the particles was governed by a magnetic field generated by a permanent magnet mounted on a servo motor. Propulsion was induced by a UV LED illuminating the sample through the objective lens. Both the servo motor and UV LED were controlled by an Arduino microcontroller.
\begin{figure}[h]
    \centering
    \includegraphics[width=0.75\linewidth]{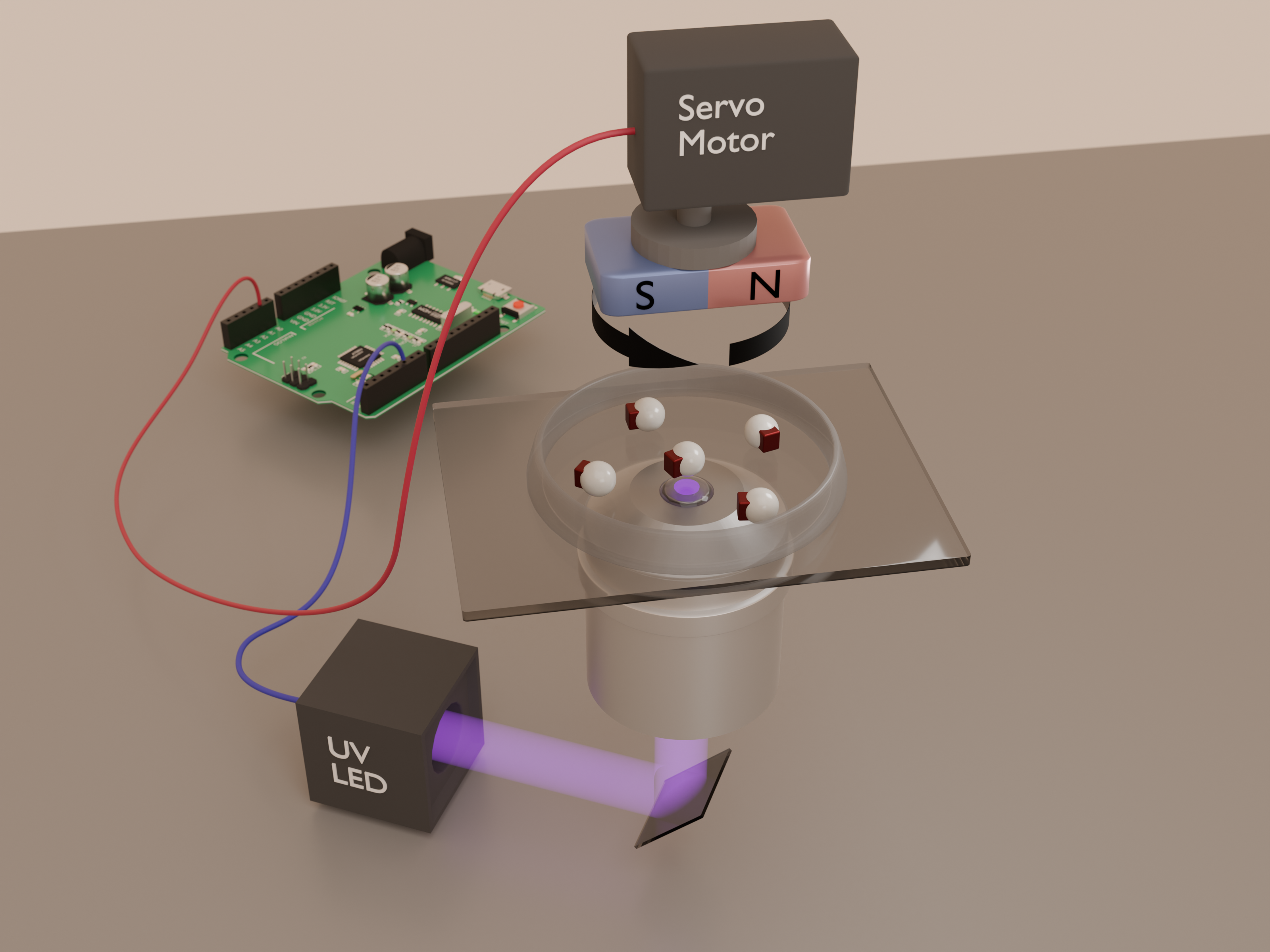}
    \caption[setup]{Schematic representation of the experimental setup used to control the motion of the particles. The Arduino Uno R3 blend was adapted from the works of unagiMOD (https://www.blendswap.com/blend/21246) released under Creative Commons Attribution 3.0.}
    \label{fig:Setup}
\end{figure}

\section*{Supplementary Note 3: Detection of turn points}\label{SM:turnpointdetection}
Experiments involving random walks were analyzed by identifying runs and turns, where a run is defined as the trajectory between two consecutive turns. Turning points were detected using different strategies tailored to each random walk type, as described below.

\subsection*{L\'evy walks}
Prior to discussing how the turning points of L\'evy were identified, it is important to mention that RTP motion, Gaussian walks, and SAWs were simulated in real-time during the experiment. The random number generation, selection of turn direction, and calculation of run-time were executed within the loop statement of the Arduino program, meaning these parameters were not predetermined.

However, implementing L\'evy walks in real-time posed challenges. The statistical properties of L\'evy walks depend strongly on long run-lengths and subtle variations in the decay of the run-time distributions for different scaling exponents ($\gamma$). Experimentally, achieving long run-lengths was constrained by the field of view, which limits the observable trajectory length.

\begin{figure}[h]
    \centering
    \includegraphics[width=0.9\textwidth]{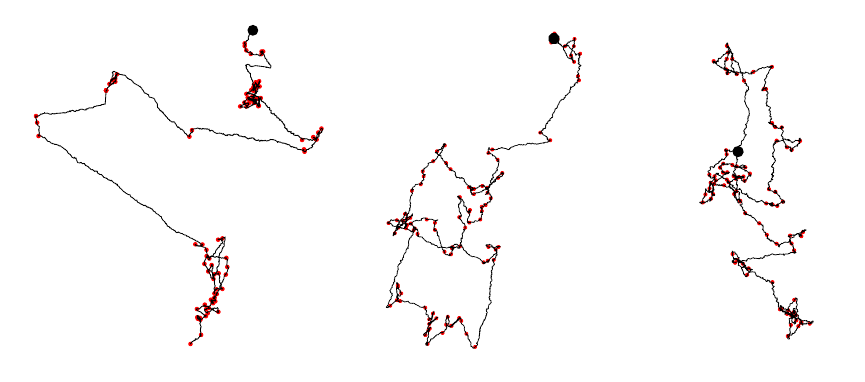}
    \caption{Experimental L\'evy walk trajectories for $\gamma$ = 1.2, 1.5, and 1.8 are shown from left to right, respectively. Detected turning points are marked as red spots along the black trajectories, with the black circle indicating the starting point of each trajectory.}
    \label{fig:turnpointdetection}
\end{figure}

The most reliable approach for implementing L\'evy walks was to pre-simulate the trajectories before the experiment. This ensured a reasonable run-length distribution while accounting for the limitations in the field of view. With pre-simulated run- and turn-times, identification of turning points in the experimental trajectories was straightforward, as illustrated in Fig.~\ref{fig:turnpointdetection}. These turning points were then used to compute the run-length distribution.

\subsection*{Run-and-tumble}
For RTP motion, turning points were identified based on background light intensity. During runs, the UV light was on, while during turns, it was switched off. The mean intensity of each frame was analyzed in Fiji (ImageJ) \cite{schindelin2012fiji,tinevez2017trackmate} to determine the timing and duration of turns. This data was then used to extract the run-time and run-length.

\subsection*{Gaussian walks}
For Gaussian walks, where the UV light intensity was constant, turns were detected based on the instantaneous angular rotation of the particles. During runs, the particle rotates very little, while during a turn, it rotates significantly. Tracked particle trajectories were used to calculate the instantaneous angular rotation (the rotation of the particle between two successive frames), which was filtered using a low-pass filter to remove high-frequency noise. Turns were identified by applying a peak detection algorithm on the turn angles with an appropriate lower bound threshold to exclude false turns.

\subsection*{Self-avoiding walks}
For SAWs, the run-time was fixed at 5\,s, making the identification of turning points straightforward once an initial reference point was established.

\section*{Supplementary Note 4: Velocity statistics}

\begin{figure}[h]
    \centering
    \includegraphics[width=0.5\linewidth]{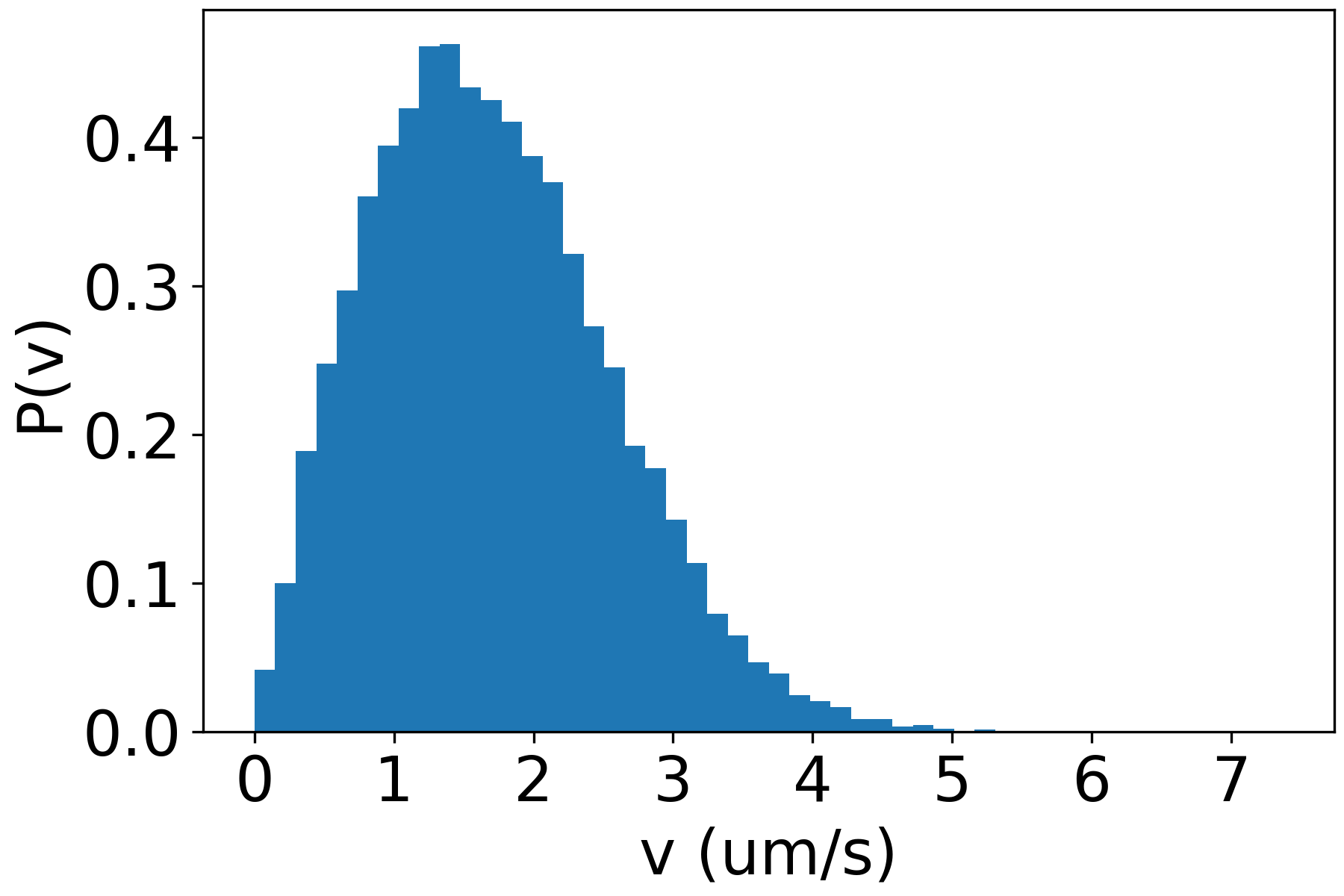}
    \caption{Probability distribution of instantaneous velocities of 9 particles over 850 s with UV light ON and magnetic field present throughout the experiment, illustrating the stochasticity in the instantaneous velocity originating from the propulsion reaction and thermal fluctuations.}
    \label{fig:veldist}
\end{figure}

Since the propulsion of our particle is generated by a chemical reaction, which is stochastic in nature, the propulsion velocities of the particles also become stochastic, as evident in Fig. \ref{fig:veldist}.

\begin{figure}[H]
    \centering
    \includegraphics[width=0.5\linewidth]{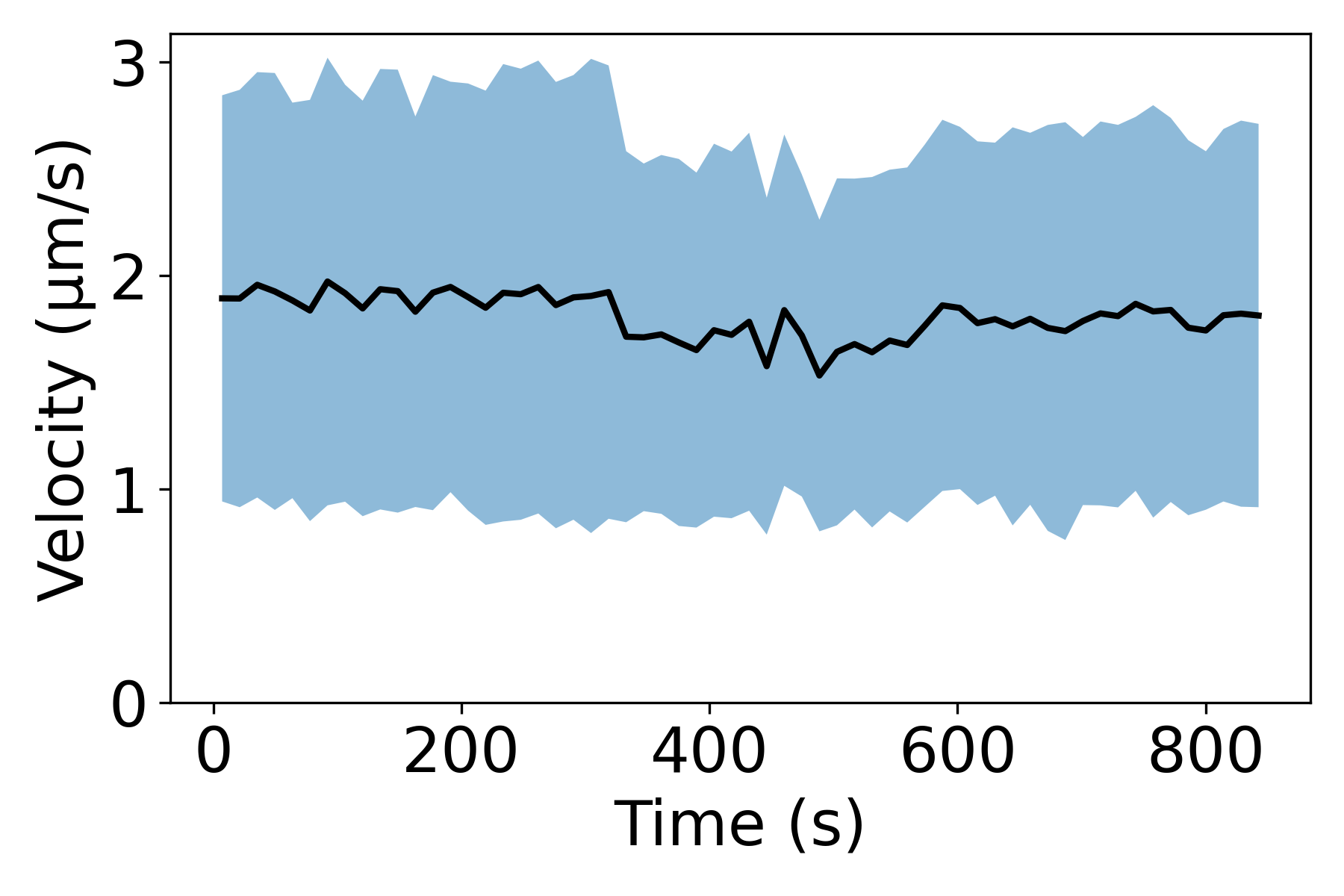}
    \caption{Average instantaneous velocity vs time in our longest Levy Walk experiment ($\gamma=1.2$). The black line represents the average over 9 trajectories with a bin size of $14\, s$, and the blue-shaded region indicates the standard deviation.}
    \label{fig:vvst}
\end{figure}

Furthermore, we also examine the effect of fuel depletion by analyzing the average instantaneous velocity over time in our longest experiment. Fig. \ref{fig:vvst} reveals that there is no significant variation in the average velocity due to fuel depletion in our system.

\section*{Supplementary Note 5: VACF of Levy Walks}
The velocity autocorrelation function (VACF) of Levy Walks with different $\gamma$ values was calculated using Eq. \ref{eq:VACf} from trajectories of the particles extracted using the Fiji plugin TrackMate \cite{tinevez2017trackmate}. VACF is a measure of persistence of motion that quantifies how correlated the velocity of a particle is with its own velocity after a time interval $\Delta t$, given by:

\begin{equation}
    C_{\mathrm{v}}(\Delta t) = \biggl \langle \frac{\langle \mathbf{v}_i(t+\Delta t) \cdot \mathbf{v}_i(t) \rangle_{t}}{\langle \mathbf{v}_i(t) \cdot \mathbf{v}_i(t) \rangle_{t}} \biggr \rangle _i
    \label{eq:VACf}
\end{equation}
where $\mathbf{v}_i(t)$ is the velocity vector of the i-th particle at time t, $\langle\cdot\rangle_i$ is the average over all particles i, and $\langle\cdot\rangle_t$ is the average over all time intervals. For a Levy Walk, the VACF is expected to show a power law decay \cite{karani2019tuning}, given by:

\[
C_{\mathrm{v}}(t) = \frac{v^{2}}{(\langle\tau_\mathrm{R} \rangle + \tau_\mathrm{T})}
\left[ \frac{\tau_{0}^{\gamma}}{(\gamma - 1)} t^{\,1-\gamma} \right]
\qquad \text{for } t > \tau_{0}
\]
where $\langle\tau_\mathrm{R} \rangle$ is the average run time, $\tau_\mathrm{T}$ is the turn time and $\tau_{0}$ is the lower cutoff for the run times.

\begin{figure}[H]
    \centering
    \includegraphics[width=0.6\linewidth]{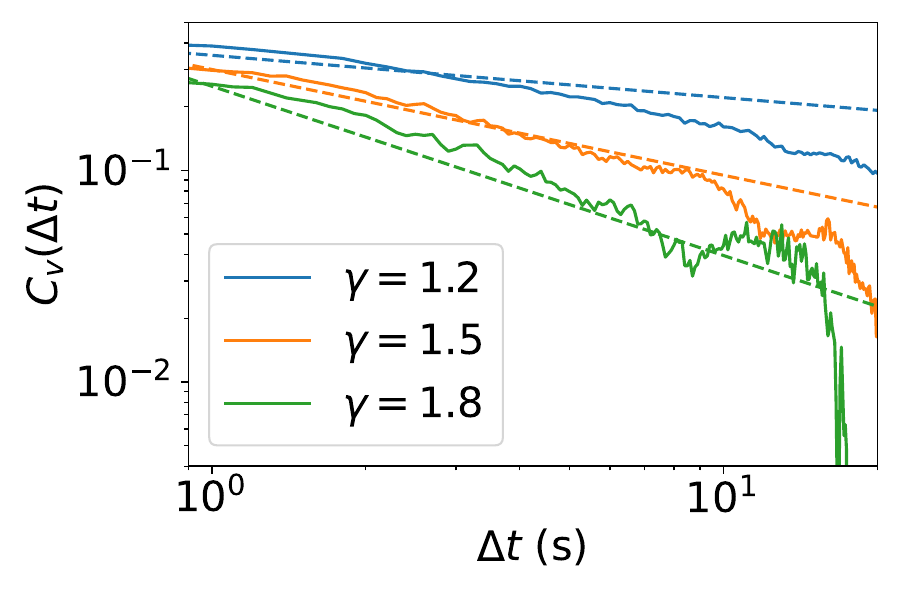}
    \caption{Velocity Autocorrelation Function $C_{\mathrm{v}}(\Delta t)$ of L\'evy Walks with different $\gamma$. The dashed lines show the expected theoretical power law decay with exponent close to $1-\gamma$ \cite{karani2019tuning}. For $\gamma=1.5 \text{ and }1.8$, the profile is close to the expected power-law decay, while $\gamma=1.2$ shows the largest deviation due to limited statistics.}
    \label{fig:placeholder}
\end{figure}

\section*{Supplementary Note 6: Real-time control of magnetic field orientation}\label{SM:joystickcontrol}
\begin{figure}[H]
    \centering
    \includegraphics[width=1\linewidth]{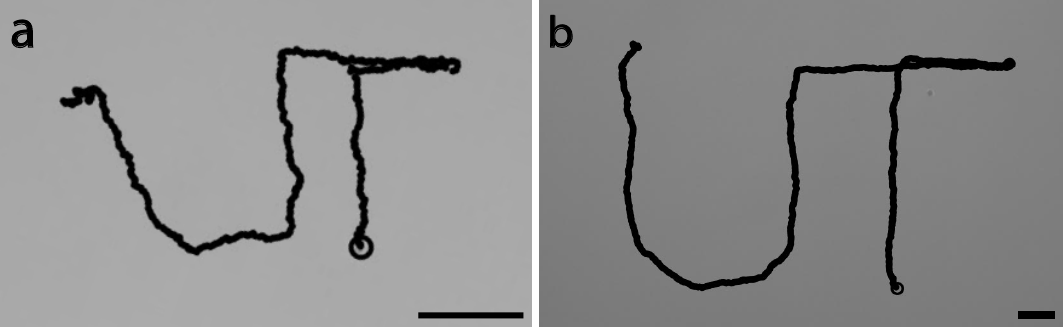}
    \caption{Comparison of pre-programmed (a) and real-time controlled (b) particle trajectories. Scale bars are 10.0\,\textmu m.}
    \label{fig:preprogrammed_vs_realtime}
\end{figure}

Arduino can be used not only to encode specific movement patterns through loop statements but also to provide real-time feedback \cite{mano2017optimal} for controlling the rotating magnet with a joystick. The joystick’s extreme positions were mapped to the maximum counterclockwise and clockwise rotation speeds, enabling continuous adjustment of the magnet’s orientation. Figure~\ref{fig:preprogrammed_vs_realtime} compares pre-programmed and real-time controlled trajectories. In the pre-programmed case, thermal fluctuations led to deviations from the intended path, whereas real-time control allowed for immediate corrections, resulting in a more precise trajectory.

We emphasize that both methods offer distinct advantages: the pre-programmed approach is more appropriate for stochastic applications, such as simulating space exploration using random walks, particularly when the trajectory or target is unknown. On the other hand, real-time control is more effective when the path is predetermined, allowing for precise adjustments during the experiment.

\section*{Supplementary Note 7: MSD and MSAD of circular swimmers}

\begin{figure}[H]
    \centering
    \includegraphics[width=\linewidth]{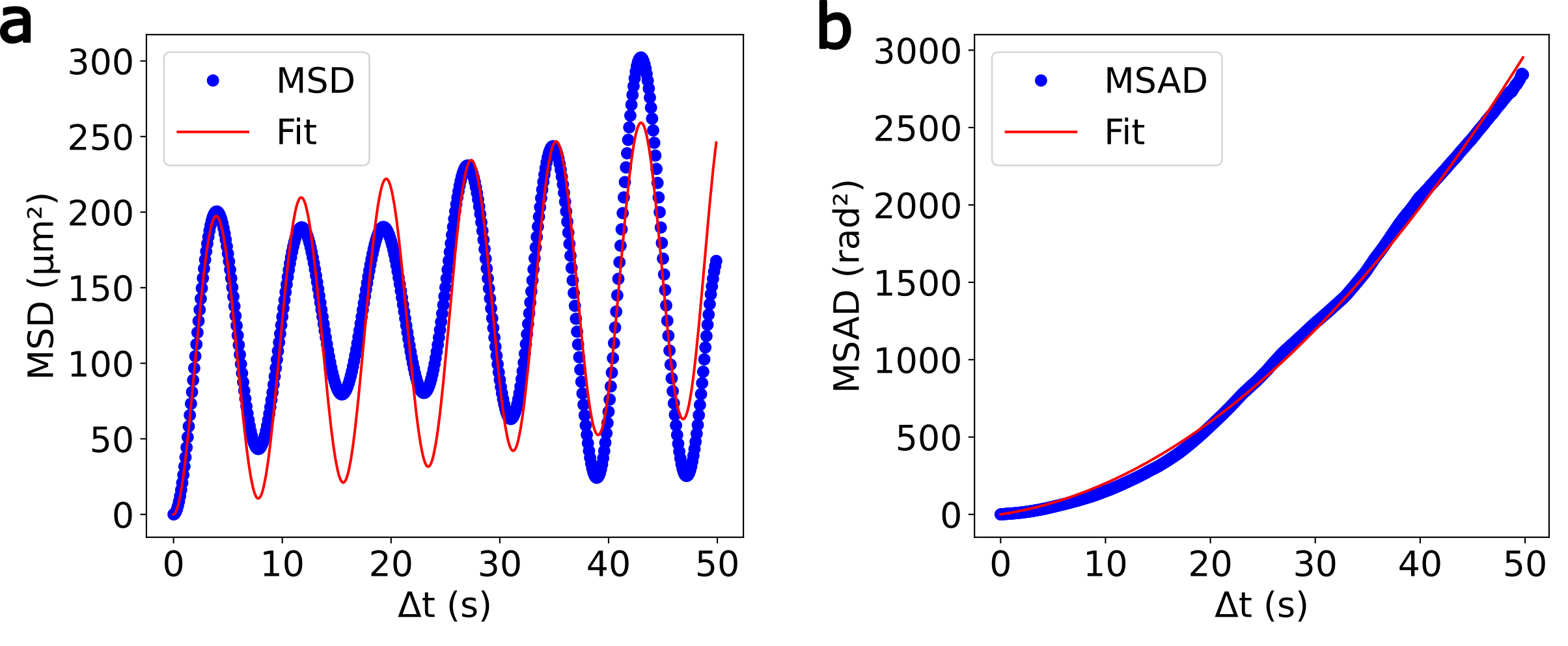}
    \caption{(a) Mean squared displacement of the circular swimmers in Fig. 5a. (b) Mean squared angular displacement of the same circular swimmers. Both plots are averaged over 2 trajectories. Red lines show fits with the expected equations for circular swimmers.}
    \label{fig:circle_msd}
\end{figure}

To further probe the single-particle dynamics of chiral active particles, we analyze the mean-squared displacement and angular displacement of the circular swimmers in Fig. \ref{fig:circle_msd}. The mean-squared displacement of circular swimmers exhibits non-monotonic behavior with oscillations, distinct from the other propulsion modes investigated. Whereas the angular displacement is monotonic with $\omega = 0.8$ rad s\textsuperscript{-1}. The MSD and MSAD fit reasonably well with the following expected equations for chiral(circular) swimmers \cite{vutukuriRationalDesignDynamics2017}.

\begin{equation}
\langle \Delta\theta^2 \rangle
= \big\langle [\theta(t) - \theta(0)]^2 \big\rangle
= \omega^2 t^2 + 2D_\mathrm{r} t
\end{equation}

\begin{multline}
\langle \Delta L^2 \rangle
=
\frac{2v^2}{(D_\mathrm{r}^2 + \omega^2)^2}
\{
(D_\mathrm{r}^2 + \omega^2)D_\mathrm{r} t
+ (\omega^2 - D_\mathrm{r}^2)\\
+ e^{-D_\mathrm{r} t} \left( (D_\mathrm{r}^2 - \omega^2)\cos(\omega t) - 2\omega D_\mathrm{r} \sin(\omega t) \right) \}+ 4 D_\mathrm{t}t
\end{multline}

\section*{Supplementary Note 8: Cluster size in collective dynamics}

Clusters were identified using the DBSCAN algorithm from the scikit-learn library\cite{scikit-learn}, with a radial distance cutoff of $1.5\,\mu$m and \texttt{min\_samples} = 2. These parameters allowed for the reliable detection of clusters. The average cluster size at time t is calculated from the cluster size (defined as the number of particles in a cluster) using:
\[C(t) = \sum_{n=0}^N n P(n,t)\]
where $P(n,t)$ is the probability of finding a cluster of size n at time t.

\begin{figure}[H]
    \centering
    \includegraphics[width=0.75\linewidth]{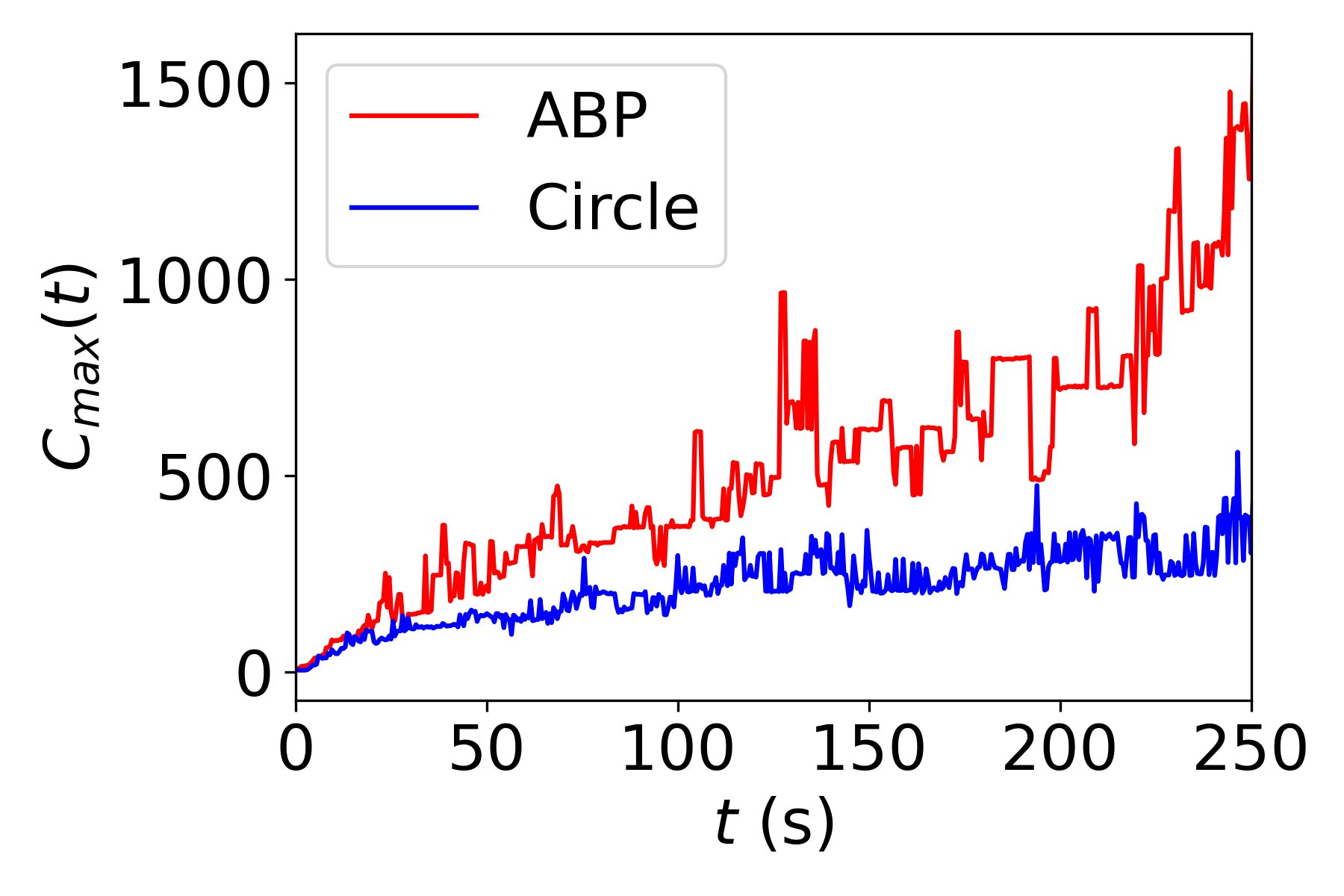}
    \caption{Number of particles in the largest cluster, $C_\mathrm{max}(t)$, as a function of time for clustering dynamics experiments of particles following circular trajectories (blue) and ABPs (red).}
    \label{fig:maxclust}
\end{figure}

While both ABPs and circular swimmers show continuous growth in the average cluster size because the smaller clusters keep growing in both cases, clusters formed by circular swimmers do not grow beyond a critical size, as reflected in the size of the largest cluster in Fig. \ref{fig:maxclust}.

\end{document}